\documentclass[twocolumn,amsmath,amssymb,amsthm,PRE]{revtex4}
\usepackage[latin1]{inputenc}
\usepackage{mathrsfs}
\usepackage{graphicx,epsf,subfigure,pstricks}
\usepackage{color}
\usepackage{float}
\usepackage{pdfsync}
\begin{document}

\author{
Thomas Cambau$^1$
}
\author{
J\'er\'emy Hure$^1$
}
\author{Jo\"el Marthelot$^{1,2}$
}

\affiliation{$^1$ PMMH, CNRS UMR 7636, UPMC \& Univ. Paris Diderot, ESPCI-ParisTech, 10 rue Vauquelin, 75231 Paris Cedex 05, France.\\$^2$ SVI, CNRS UMR 125, Saint-Gobain, 39 Quai Lucien Lefranc, 93303 Aubervilliers Cedex, France.}

\title{Local stresses in the Janssen granular column}

\begin{abstract}
We study experimentally the distribution of local stresses in a granular material confined inside a vertical cylinder. We use an image correlation technique to measure the displacement field of the container induced by the forces exerted by the grains on the inner wall. We describe an optimization procedure based on the linear theory of elastic shells to deduce the distribution of these forces from the measured displacement field. They correspond to the stress field of the granular material close to the container's inner wall. We first confirm the validity of Janssen's description for various experiments, including the influence of the beads diameter and the effect of an additional mass on top of the granular column. We then apply this method to determine the stress field during the gravity driven discharge of a silo through an aperture.  
\end{abstract}

\maketitle 

\section{Introduction}
The stress distribution in a granular material is a complex problem illustrated by two canonical experiments: the sandpile and the silo. In 1829, Huber-Burnand \cite{huber} noticed that an egg covered with several inches of sand was able to support a mass of iron weighing fifty-five pounds without breaking, prefigurating the studies of forces repartition in granular materials \cite{smid81}. The importance of the construction history on stress distributions under sandpiles was then demonstrated \cite{Vanel99}. In 1895, Janssen \cite{sperl2006} quantified the saturation of the bottom pressure in a granular material confined in a vertical container. The so-called \textit{Janssen's law} defines the typical pressure saturation length, accounting for friction of the grains on the container's wall. 

The practical interest in avoiding the damage and collapse of silos \cite{dogangun,zhong} has motivated numerous studies. More recently, the confined granular column has been extensively studied as the simplest experimental setup to test the influence of many parameters on the repartition of stresses in granular materials. The hypothesis of fully mobilized friction \cite{evesque1998,pgg}, the effects of humidity \cite{bertho2004} and the motion of the wall of the silo \cite{bertho2003} have been assessed through careful experiments. Such experiments have been used as tests to validate continuous models of granular materials, as oriented stress linearity (OSL) \cite{vanel99b,vanel2000} or elastic \cite{ovarlez1,ovarlez2, goldenberg05} theories.

In all these studies on confined granular materials, stresses all along the container's wall were inferred from the measurement of the apparent mass at the bottom of the column. The total shear stress on the column was obtained directly by measuring the mass of the lateral wall \cite{perge}. Apart from numerical results \cite{landry2003}, direct measurements of local stresses remain scarce. In a recent work, the local force network in a 2D silo was  directly measured with photoelastic particles \cite{Wambaugh2010}. The conclusions from these experiments are puzzling, showing significant deviations from Janssen-like models. This clearly indicates that direct measurements of stresses for static confined granular material are still needed.
 
We propose here a different method to directly measure the stresses at the wall by tracking the minute deformation of the silo. We follow an idea proposed by Janssen in his original paper: \textit {it was the intention of the author to determine the side pressure of the corn directly in the experiments} by using a side lid pushed against the silo's walls. However in his experimental setup, he noticed that \textit{accurate results could not be obtained} \cite{sperl2006}. 

In addition to the static Janssen's law, the dynamics of confined granular material have also been studied extensively. The flow rate of the gravity driven discharge of a granular column in a silo is described by \textit{Beverloo's law} \cite{beverloo61}. Even though well verified experimentally \cite{Mankoc07}, the physical meaning of this Beverloo's law, sometimes described as a consequence of Janssen's law \cite{andreotti}, remains unclear. The recent measurements of the pressure profile in the outlet plane of a discharging silo have shown that the flow-rates are not controlled by the local stress conditions \cite{perge}. The local displacement field of the grains exhibits complex patterns, such as shear zones close to the wall \cite{Pouliquen,Midi04}. Diffusive-wave spectroscopy experiments show the existence of slow collective rearrangements in addition to fast grain collisions \cite{Menon97}. The friction mobilization at the silo's wall during the discharge must also be described to understand collapse events \cite{guti}. Measurements of local stresses in confined granular materials are thus also required to gain insight into the physics of Beverloo's law.

The paper is organized as follows. In Section~II, we recover Janssen's law from mass measurement at the bottom of silos with different elastic moduli. For the given set of parameters used in this study, there is no dependence of the saturation mass with the mechanical properties of the container. This leads to an experimental setup to measure wall deformations and infer the stress distribution in the granular material close to the wall, detailed in Section~III. The technique is validated against indirect results for quasi-static experiments in Section~IV-A. We finally use this method to gain insight into the evolution of local stresses distribution during the discharge of granular matter through an aperture.
\section{Janssen's law in a soft container}
\label{sec2}
\subsection{Material and setup}

\begin{figure}[!h]
\centering
  \includegraphics[height=6cm]{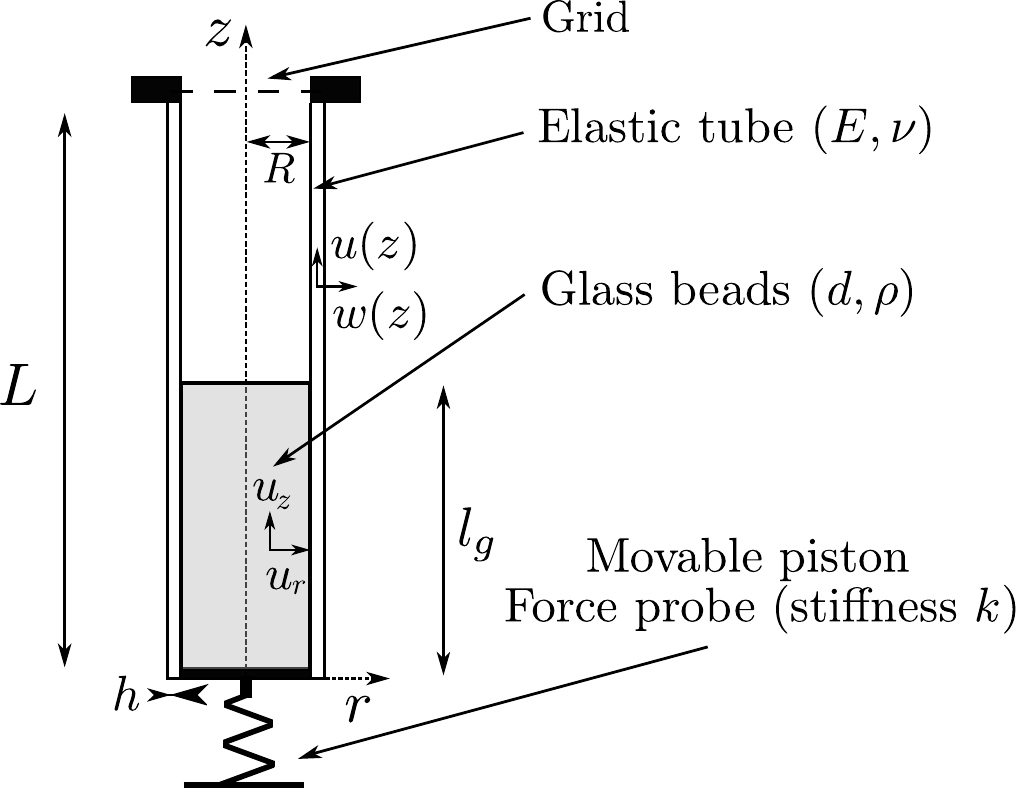}
  \caption{Experimental setup. A tube (length $L$, radius $R$, thickness $h$, Young's modulus $E$ and Poisson's ratio $\nu$) clamped at its top is rain filled with glass beads (diameter $d$ and density $\rho$), up to a height $z = l_g$. The bottom of the tube is closed by a piston connected to a force probe of stiffness $k$. The radius of the piston is slightly lower than $R$ to prevent friction. The radial and longitudinal displacements of the granular material and the tube are noted $\{u_r,u_z\}$ and $\{u,w\}$, respectively.}
  \label{fig1}
\end{figure}

Experiments were performed in elastomer cylindrical tubes and compared to a reference experiment in a rigid tube made of Poly(methyl methacrylate) (PMMA). We used silos of radius $R = 18\,$mm, thickness $h = 2\,$mm made of shore 8 (Young's modulus $E = 0.24\,$MPa) and shore 32 ($E = 0.96\,$MPa) PolyVinylSiloxane (PVS) or PMMA ($E = 2.5\,$GPa). The tubes are clamped at the top and free to move at the bottom (Fig.~\ref{fig1}). The granular material stands on a movable piston connected to a force probe to measure the apparent mass at the bottom of the silo. The grains are dry, non-cohesive and slightly polydisperse ($10\%$) glass beads of diameter $d = 1.5\,$mm (unless otherwise specified) and density $\rho =  2545\,\mathrm{kg.m^{-3}}$. The PVS tubes are dusted with talc powder to prevent any adhesion between the grains and the wall of the silo. The static friction coefficient $\mu_s$ between the glass beads and the PMMA and PVS was measured using the sliding angle of a three-bead tripod. We found $\mu_s = 0.5 \pm 0.1$ for the PMMA tube, $\mu_s = 0.51\pm0.04$ (resp. $\mu_s = 0.52\pm0.03$) for the shore 32 (resp. shore 8) PVS tube. In the following, the static friction coefficients are thus assumed to be the same in all experiments. The dynamic friction coefficient was measured with the same apparatus : once set into motion, the angle is slightly reduced until the three-bead tripod eventually stops. This leads to $\mu_d = 0.43 \pm 0.03$ for the shore 8 PVS tube. All the experiments were performed at room temperature and at a relative humidity of $35\pm5\%$ at least twice to ensure repeatibility.\\

\subsection{Saturation mass at the bottom of the silo.}
\begin{figure}[h]
\centering
  \includegraphics[height=5.8cm]{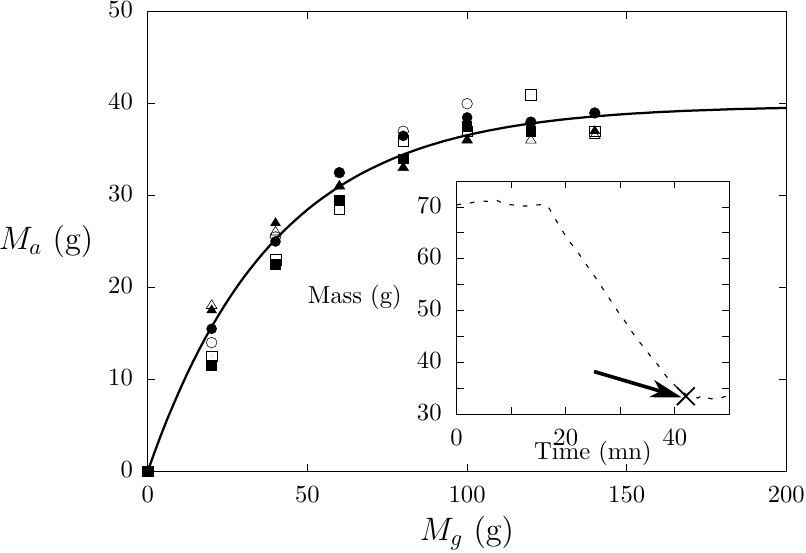}
  \caption{Evolution of the apparent mass $M_a$ as a function of the filling mass $M_g$. $M_a$ was measured by the force probe at the bottom of the silo as in \cite{vanel2000,ovarlez1}. Triangles, squares and circles respectively correspond to the PVS shore 8 ($E=0.24$ MPa), PVS shore 32 ($0.96$ MPa) and PMMA ($2.5$ GPa) tubes. Each experiment was performed twice (open and filled symbols). The solid line corresponds to Janssen's model $M_a  = M_{sat}[1 - \exp{(-M_g/M_{sat})]}$, with $M_{sat} = 40$ g best fitting the experimental data. Inset: Evolution of the measured mass as a function of time as the piston is lowered ($M_g = 80\,$g in PVS shore 32). The cross represents the measured point $M_a$ which corresponds to a fully mobilized friction and provides a single data point in the main graph.}
\label{fig3}
\end{figure}

A mass $M_g$ of beads is poured through a $5\,$mm grid. Rain filling provides reproducible dense piling (with volume fraction $\phi \approx 64\%$). The piston is then moved downwards at constant speed $V = 1.5\,\mu \mathrm{m.s^{-1}}$, to mobilize the friction of the grains on the wall. The apparent mass is measured as a function of time (Inset Fig.~\ref{fig3}) and decreases until it reaches a plateau, noted $M_a$, when the friction at the wall is at the Coulomb threshold as described in \cite{vanel2000,ovarlez1}. We plot $M_a$ as a function of the mass of beads filling the tube (Fig.~\ref{fig3}). The apparent mass $M_a$ increases with $M_g$ until reaching a plateau noted $M_{sat}$.\\ 
 
At equilibrium, three forces resist the weight $-\rho \phi g \pi R^2 dz$ of a slice of height $dz$, where $g$ is the gravitational acceleration. The upper part of the granular material applies a force $\sigma_{zz}(z+dz) \pi R^2$, the lower part  $-\sigma_{zz}(z) \pi R^2$, and friction on the wall $\sigma_{rz}(z) 2 \pi R dz$, where $\sigma_{rr }$,  $\sigma_{zz }$ and  $\sigma_{rz }$ are average stresses in the grains on the radial direction. Assuming that the shear stress follows Coulomb law at the threshold, wall friction can be rewritten as $- \mu_s \sigma_{rr}(z) 2 \pi R dz$. Finally, Janssen assumed a constant redistribution of stresses $\sigma_{rr} = K \sigma_{zz}$, leading to the equilibrium equation $d\sigma_{zz}/dz - 2K\mu_s \sigma_{zz}/R = \rho \phi g$. The solution is :
\begin{equation}
	\sigma_{zz} = -\frac{\rho \phi g R}{2 K \mu_s} \left[1 - \exp \left( \frac{2 K \mu_s (z-l_g)}{R} \right) \right]
\end{equation}
where $z = 0$ is the bottom of the tube. The apparent mass $M_a = [\pi R^2/g] |\sigma_{zz} (z=0)|$ at the bottom of the tube is finally :
\begin{equation}
M_a = M_{sat}\left[1 - \exp{\left(- \frac{M_g}{M_{sat}}\right)}\right]
\end{equation}
where $M_{sat} = (\pi R^3 \rho \phi /2 K \mu_s)$ is the measured saturation mass and $M_g = \pi R^2 l_g \rho \phi$ is the mass of grains filling the tube. Fitting the experimental data with Janssen's model (solid line Fig.~\ref{fig3}), we find $M_{sat}=39.8$ g, leading to $K= (\pi R^3 \rho \phi)/(2 M_{sat} \mu_s) =0.75$.\\

The evolution of the apparent mass as a function of the mass of beads filling the tube follows Janssen's law and, for our experimental parameters, does not depend on the elasticity of the silo. Indeed, the only quantity that can be changed in Janssen's model accounting for the elasticity of the silo is the radius $R$ resulting from the pressure of the grains on the tube. Therefore, in the limit of small deflections, $w/R \ll 1$, the stress distribution will not be affected. More quantitatively, the pressure applied on the wall scales as $|p| = \sigma_{rr} \sim \rho g R$. The radial displacement of an elastic tube subjected to this pressure can be written as $w \sim pR^2 / Eh \sim \rho g R^3 /Eh$ \cite{timo}. Thus the ratio $w/R$ scales as $\rho g R^2/Eh$, which is less than $10^{-2}$ in our experiment. Previous studies have shown that the apparent mass is very sensitive to the compaction \cite{vanel99b}, thus the radial displacement of the tube has to be small enough not to change the initial compaction. This can be written as $w < d$, where $d$ is the diameter of the glass beads.\\

Before describing the method to measure local stresses in the granular material, we compare these experimental results to the prediction of a model that assumes isotropic elasticity for the grains.

\subsection{Isotropic elasticity model}

Assuming isotropic elasticity for the granular material, with Young's modulus $E_g$ and Poisson's ration $\nu_g$, Ovarlez \textit{et. al.} \cite{ovarlez2} showed numerically a strong dependence of the saturation mass with the ratio $E_g/E$. More quantitatively, we can follow the analytical description given in \cite{ovarlez2}, adding the elasticity of the tube. Far from the free surface and from the piston, both stresses and strains of the granular material should be independent of the coordinate $z$. The non-zero terms of the strain tensor $\epsilon$ can be written as a function of the radial $u_r$ and longitudinal $u_z$ displacements:\begin{equation}
\epsilon_{rr} = \frac{\partial u_r}{\partial r} \ \ \ \ \ \epsilon_{\theta \theta} = \frac{u_r}{r} \ \ \ \ \ \epsilon_{zz} = \frac{\partial u_z}{\partial z} \ \ \ \ \ \epsilon_{r z} = \frac{1}{2} \frac{\partial u_z}{\partial r}
\end{equation}
In this asymptotic regime, the equilibrium equations
\begin{equation}
\frac{\partial \sigma_{rr}}{\partial r} + \frac{\sigma_{rr} - \sigma_{\theta \theta}}{r} = 0 \ \ \ \ \ \ \ \ \frac{\partial \sigma_{rz}}{\partial r} + \frac{\sigma_{rz}}{r} = \rho g
\label{equielas2}
\end{equation}
are solved by assuming Hooke's law for the granular material:
\begin{equation}
\epsilon_{\alpha \beta } = \frac{1+\nu_g}{E_g} \sigma_{\alpha \beta} - \nu_g \delta_{\alpha \beta} \sigma_{\gamma \gamma}
\end{equation}
where $(\alpha,\beta,\gamma) \in \{r,\theta,z \}$, and $\delta$ is the Kronecker symbol, and with the boudary conditions $\sigma_{rz}(r=R,z) = -\mu_s \sigma_{rr}(r=R,z)$ This yields the radial and longitudinal displacements:
\begin{subequations}
\begin{align}
u_r(r,z) &= \frac{U_R}{R}\,r \\ 
u_z(r,z) &= \frac{1+\nu_g}{2E_g} \rho g r^2 - z \left(\frac{(1-2\nu_g)(1+\nu_g)}{\nu_gE_g } \frac{\rho g R}{2\mu_s} + \frac{1}{\nu_g}\frac{U_R}{R}  \right)
\end{align}
\label{ovarlezuruz}
\end{subequations}
with $U_R$ the radial displacement of the tube. Setting $U_R$ to $0$ in Eqs.~\ref{ovarlezuruz}, we recover the displacements given in \cite{ovarlez2}. The displacement $U_R$ is related to the pressure $p = -\sigma_{rr}=(\rho g R)/(2 \mu_s)$ acting on the tube: 
\begin{equation}
U_R = \frac{p R^2}{Eh}=\frac{\rho g R^3}{2\mu_s Eh}
\end{equation}
Finally, the saturation mass $M_{sat,elas} = \left(\pi R^2/g \right) |\sigma_{zz}|$ takes the form:
\begin{equation}
M_{sat,elas} = M_{sat} \left( 1 + \frac{E_g}{(1-\nu_g)E}\frac{R}{h}\right)
\label{msatelas}
\end{equation}
where $M_{sat}$ is the saturation mass considering a rigid tube. Experimentally, we varied the ratio $E_g/E$ by a factor $10^4$, and did not see any evidence of this dependence. The isotropic elasticity description is thus insufficient to describe confined granular material and we need to consider anisotropy or elliptic theories, as pointed out in \cite{ovarlez2}. Nevertheless, the stress distribution in the granular material appears to be the same inside a rigid or soft container, which leads to the method for measuring local stresses detailed in the next section. \\

\section{Evaluation of local stresses}
\subsection{Experimental setup}
The saturation of pressure with depth in a granular column is a consequence of friction of the granular material on the silo's inner wall. We propose to measure the resulting displacement field of the outer wall of the silo to reconstruct the stress field of the granular material close to the wall. Although difficult to estimate for industrial silos \cite{zhong}, we will show that such displacements may be obtained by using soft elastomer containers. As shown in Fig.~\ref{fig3}, there is no perceptible effect of the elasticity of the silo in our experiments, we thus use the shore 8 tube in the following.\\

\begin{figure}[h]
\centering
  \includegraphics[height=5cm]{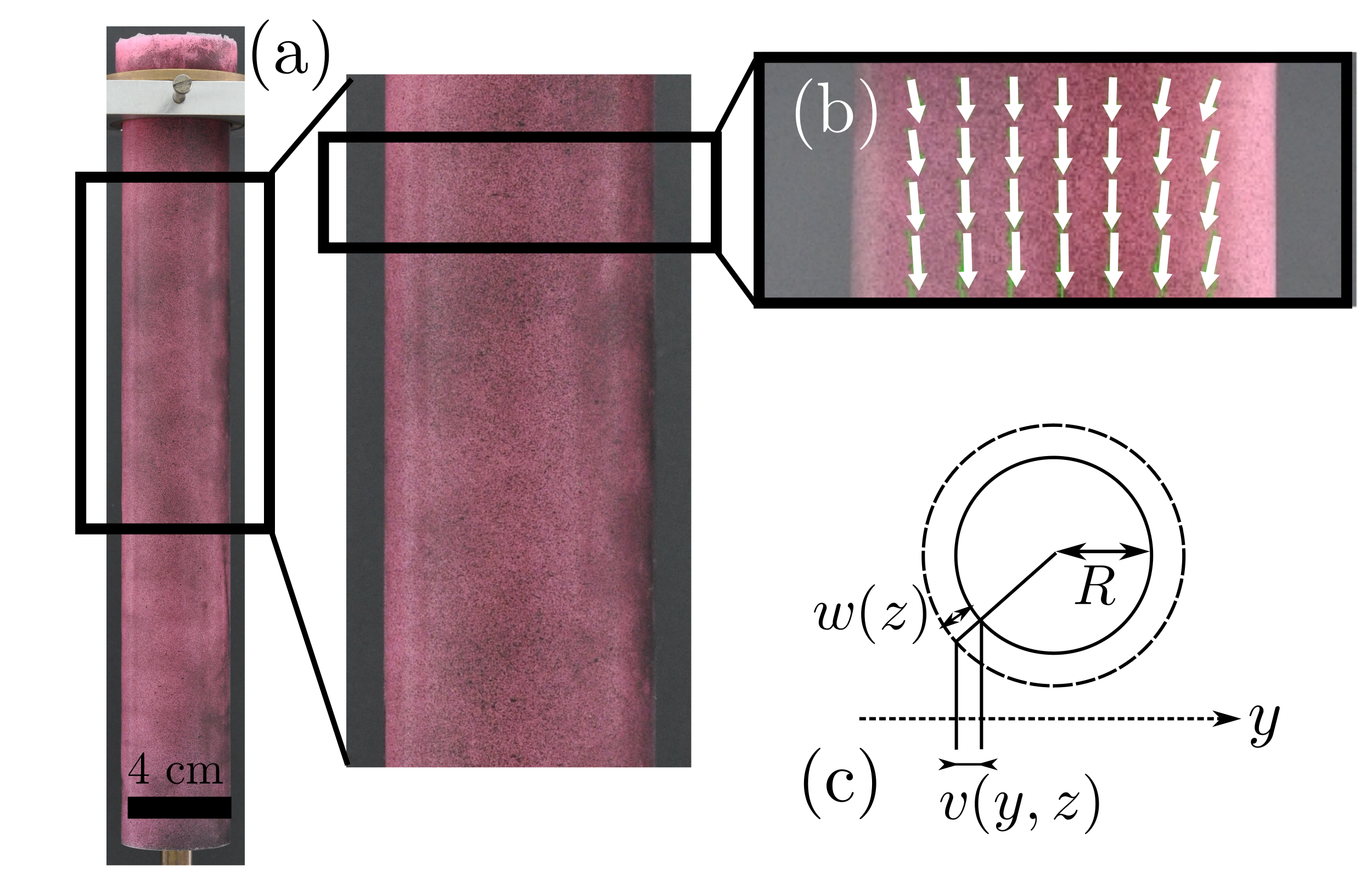}
  \caption{(Color online) (a,b) The PVS tube is sputtered with black paint to measure the displacement field (arrows) using an image correlation technique (PIVlab \protect\cite{pivlab} with Matlab). A reference picture is taken when the tube is empty. This image is then compared to another one obtained when the apparent mass saturates (quasi-static experiments), or at a given time $\Delta t$ after the plug was removed (discharge experiments). (c) Image correlation gives both the longitudinal displacement $u(y,z)$ and the transverse displacement $v(y,z)$. The radial displacement can be obtained following $v(y,z) = (y/R)\,w(z)$.}
\label{fig2}
\end{figure}

\vspace{0.5cm}
\noindent
\textit{Quasi-static experiments} --
We use the procedure described in Section II.B. The PVS tubes are additionally sputtered with black paint, as shown on Fig.~\ref{fig2}. Displacements were obtained by correlating a picture of the tube, taken as the apparent mass reaches the Janssen's plateau, with a reference picture of the empty tube \protect\footnote{Pictures were taken using a Nikon D200 SLR camera, with a resolution of 4288x2848 pixels. A pixel size is about 0.05 mm}. The cross-correlation of the two pictures (PIVlab \cite{pivlab} with Matlab) leads to the displacement field $\{v(y,z),u(y,z)\}$ defined on Fig.~\ref{fig1} and Fig.~\ref{fig2}. These displacement fields are finally converted into the radial and longitudinal components of the displacement $v(y,z) = (y/R)\,w(z)$ so that $w(z) = R\,dv(y,z)/dy$ and $u(z) = {\langle u(y,z) \rangle}_y$, respectively.\\
\textit{Discharge experiments} -- The piston is replaced by a fixed cylinder of radius $R_c = 17.75$ mm slightly smaller than the radius of the tube to avoid friction. A conical hole of minimal diameter $D = 11$ mm, maximal diameter $D = 22$ mm and angle $60^{\circ}$ drilled into the cylinder is closed by a plug (see Inset Fig.~\ref{fig9}). Once the tube is rain filled with grains, the plug is removed and the mass of grains flowing outside the tube is recorded as a function of time. The displacement field is obtained by performing the cross-correlation between a reference picture of the empty tube and pictures taken at a given time after releasing the plug. The displacement field $\{u(z),w(z)\}$ is obtained as in the quasi-static experiment.\\

A typical result of the image correlation is shown on Fig.~4(a,b). The tube is stretched axially as the grains pull on the wall. The radial displacement is positive in the filled region because of the pressure of the grains, and negative above because of Poisson's effect (the tube being clamped at the top). The radial and longitudinal displacements are direct consequences of the shear stress $\sigma_{rz}(R,z)$ and radial stress $\sigma_{rr}(R,z)$ inside the grains, close to the wall. In the following, we will note $\sigma_{rz}(R,z) = \sigma_{rz}(z)$ and $\sigma_{rr}(R,z) = \sigma_{rr}(z)$ to ease legibility, keeping in mind that these quantities are local and not averaged like in the Janssen model. A refined Janssen model accounting for the radial dependence leads to the same qualitative behavior \cite{rahmoun2008}.
We recall in the next paragraph the linear theory of cylindrical elastic shells, which link the applied forces on the container's inner wall to the displacement field. 

\subsection{A model for the deformation of the silo}
\subsubsection{Cylindrical shells equations}
We use the linear theory of elastic cylindrical shells \cite{timo} to describe the deformation of the silo \footnote{In the thin shell limit $h/R \ll 1$, and for axisymetric deformations, the non-linear term in axial strain  $(dw/dx)^2$ can be neglected in the limit $u/L \gg (w/L)^2$, which is the case in our experiments}. For axisymmetric deformations, the non-zero values of the strain $\epsilon$ and curvature $\kappa$ tensors are written in terms of axial $u$ and radial $w$ displacement:
\begin{equation}
\epsilon_{zz} = \frac{du}{dz} \ \ \ \ \ \ \ \ \epsilon_{\theta \theta} = \frac{w}{R} \ \ \ \ \ \ \ \ \kappa_{zz} = -\frac{d^2 w}{d z^2}
\end{equation} 
The moment $m_{\alpha \beta}$ and in-plane stress $n_{\alpha \beta}$ are given by Hooke's law:
\begin{align}
m_{zz} = \frac{Eh^3}{12(1-\nu^2)}\kappa_{zz}\\
n_{zz} = \frac{Eh}{1-\nu^2}(\epsilon_{zz} + \nu \epsilon_{\theta \theta})\\
n_{\theta \theta} = \frac{Eh}{1-\nu^2}(\epsilon_{\theta \theta} + \nu \epsilon_{zz})
\end{align}
Finally, the in-plane and out-of-plane equilibrium equations are:
\begin{align}
\label{eq2}
\frac{d n_{zz}}{d z} = -h f_v\\
\label{eq3}
\frac{d^2m_{zz}}{dz^2} - \frac{n_{\theta \theta}}{R} = -p
\end{align}
where $p$ is the pressure acting on the interior face of the tube (positive towards the exterior) and $f_v$ the axial volume force (positive upwards). In the case of a tube filled with granular materials, the pressure is given by $p = - \sigma_{rr}(z)$. Grains also apply a shear stress $-\sigma_{rz}(z)$ on the tube which can not be accounted for directly in shells equations. However, we can note that a shear force acting on a height $dz$ of tube, $-\sigma_{rz} 2 \pi (R-h/2) dz$, is equivalent to an axial volume force $f_v h 2 \pi Rdz$, where $f_v h = -(1-h/2R)\sigma_{rz}$.

For given functions $\sigma_{rr}(z)$ and $\sigma_{rz}(z)$, solving the equilibrium equations \ref{eq2} and \ref{eq3} with appropriate boundary conditions leads to the axial and radial displacements. We do not solve the equations over the entire length of the tube, as the upper part ($x>l_g$) deformation corresponds to the stretching of an empty shell \footnote{Except close to the upper clamped boundary.}, a state completely described by $u(x>l_g) = u_0 (x/l_g)$ and $w(x>l_g) = w_0$. In the following, the boundary conditions considered are $u(l_g) = u_0$, $w(l_g) = w_0$, $w'(l_g) = 0$ and a free edge at the bottom, where $n_{zz}(0) = m_{zz}(0) = dm_{zz}/dz(0) = 0$. The values $u_0$ and $w_0$ are obtained experimentally. \\

We now describe the optimization procedure to solve the inverse problem, \textit{i.e.} once knowing the displacement field, finding  the shear stress and pressure in the granular material close to the wall.

\subsubsection{Optimization procedure}

\begin{figure*}
\centering
\includegraphics[height=4.cm]{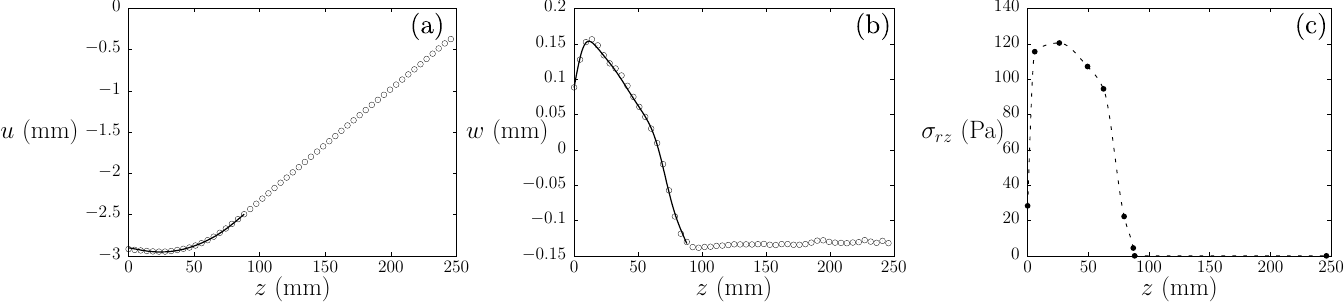}
\caption{Quasi-static experiment: Typical results for a shore 8 tube ($E = 0.24$ MPa, $\nu = 0.5$) and a filling mass $M_g=140$ g ($l_g = 80$ mm). $z=0$ corresponds to the bottom of the tube, $z=l_g$ corresponds to the free surface of the granular column. (a,b) Longitudinal and radial displacements along the tube. The circles correspond to measurements obtained by image correlation, the solid line corresponds to the result of the optimization procedure. The computation was only done in the part of the tube filled with grains, with appropriate boundary conditions to describe the empty part. (c) Shear stress in the granular material at the wall: the minimisation algorithm finds the values of the shear stress best fitting both experimental displacement curves. Shear stress is interpolated using cubic spline between the n=8 evaluation points (black dots).}
\label{fig4}
\end{figure*}

The portion of the tube filled with grains, $0 \leq z \leq l_g$, is discretized into $n$ elements. The values of the functions $\sigma_{rr}(z)$ and $\sigma_{rz}(z)$ at the nodes constitute the $2n$ unknowns of our inverse problem, and the values of the functions in each element are interpolated from these points. For a given set of unknowns, the boundary value problem (Eq.~\ref{eq2},\ref{eq3}) is solved by a collocation method (Matlab's \textit{bvp4c} solver) \cite{bvp4c}, leading to the displacement field $\{u,w\}$. The squared deviation of this field from the experimental measurement is then computed. The procedure is repeated by changing the values of the unknowns until the squared deviation is minimal, using Levenberg-Marquardt or Active-Set algorithms (Matlab's \textit{lsqnonlin} \cite{levenberg} and \textit{fmincon} \cite{fmincon} functions). The initial guess for the $2n$ unknowns is set to zero. We checked that the converged solution does not depend on the initial guess. A typical result of the optimization procedure is shown on Fig.~\ref{fig4}.

There is a unique solution to the linear system of equations (6,7). Thus in principle, increasing $n$ leads to better evaluations of the functions $\sigma_{rz}$ and $\sigma_{rr}$. However, large values of $n$ greatly increase computational cost, and are more likely to lead to local minima. Therefore, we use small values of $n$, and repeat the entire optimization procedure (typically a hundred times) for random location of the discretization nodes along the filled region of the tube. This allows to assess the values of the unknown functions everywhere along the tube in a computationally effective way. For each optimization procedure, the value of $n$ is selected as the minimal value above which the unknown functions do not change (when increasing $n$ only leads to noisy results).

\section{Experimental results}
\label{sec3}

In the first part of this section, we measure the shear stress and pressure acting on the wall in the quasi-static experiments using the optimization procedure described in the previous section. As friction is expected to be fully mobilized, we can reduce the number of unknowns by considering  $|\sigma_{rz}| = \mu |\sigma_{rr}|$, where $\mu$ is an additional unknown. Note that the optimization procedure has also been carried out for the quasi-static experiments without the hypothesis of fully mobilized friction, leading to identical results.

In a second part, we measure the evolution of stresses during discharge experiments. As the friction state is in this case unknown, we look at the values of the shear stress and pressure separately, assuming Coulomb friction law:
\begin{equation}
|\sigma_{rz}| \leq \mu |\sigma_{rr}|
\label{eqvidange}
\end{equation}
where $\mu$ is taken as the static friction coefficient $\mu_s$ measured experimentally.

\subsection{Local stresses in Janssen granular column.}
\subsubsection{Stress distribution as a function of height}
\begin{figure}[h]
\centering
\includegraphics[height=5.9cm]{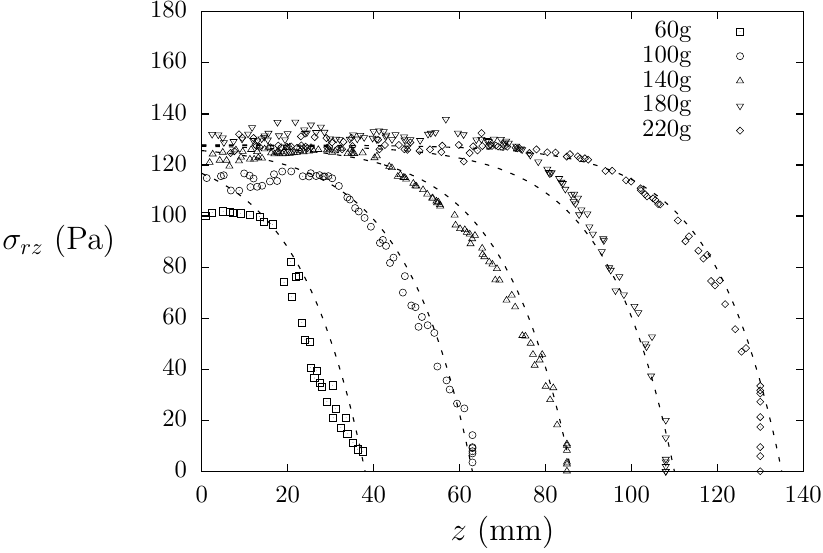}
\caption{Quasi-static experiment: Shear stress $\sigma_{rz}$ along the tube for different filling masses $M_g = 60\,$ g (squares), $100\,$ g (circles), $140\,$ g (triangles), $180\,$ g (inversed triangles), $220\,$ g (diamonds). $\sigma_{rz}$ saturates from top $z=l_g$ to bottom $z=0$. For a given experiment, the optimization procedure is repeated a hundred times with $n=5$ estimation points of the shear stress. Dashed lines correspond to a fit of the experimental data by the function $f(z)=[\mu_s/(1+2\mu_s^2)][gM_a/\pi R^2][1-\exp(K_w(z-l_g)/R)]$ (Eq.~\ref{rahmoun}), with $K_w = 1.16$.}
\label{fig5}
\end{figure}

We use the optimization procedure described above to evaluate the shear stress $\sigma_{rz}$ at the wall, for different filling masses. The shear stress $\sigma_{rz}$ increases with depth, from zero at the free surface to a saturation value (Fig.~\ref{fig5}). The saturation length is of the order of the tube diameter, independant of the filling mass and in agreement with Janssen's law. The difference between the weight of the total column $gM_g$ and the force on the wall $2\pi R \int_0^{l_g} \sigma_{rz}dz$ is in agreement with the bottom weight measurement $gM_a$. As expected, if the filling length $l_g$ is smaller ($l_g<60$ mm), stresses do not reach the saturation value. We obtained a friction coefficient $\mu = 0.46 \pm 0.04$, consistent with the static friction coefficient measured independantly.\\ 
The saturation value of the stresses can be compared to theoretical predictions. Assuming a non-cohesive material, equality of internal and beads/wall friction angles, and a Mohr-Coulomb criterion, the stress ratio at the wall can be written as  \cite{rahmoun2008}:
\begin{equation}
\lambda_w = \sigma_{rr}/\sigma_{zz} = 1/(1+2\mu_s^2)
\label{rahmoun}
\end{equation}
with $\sigma_{zz} = gM_a/\pi R^2$. This leads to $\sigma_{rr} \approx 280\,$Pa, in excellent agreement with the measured value $\sigma_{rr} = \sigma_{rz}/\mu \approx 282\,$Pa. The experimental results shown on Fig.~\ref{fig5} are in very good agreement with an exponential saturation behavior and Eq.~\ref{rahmoun} \footnote{Eq.~1 can not be used directly for comparison to our results shown in Fig.~\ref{fig5} and \ref{fig6}, as stresses in the Janssen's model are averaged along the radial direction. We thus used the exponential behavior with a parameter $K_w$, and Eq.~\ref{rahmoun} to fit our experimental results.}. Measuring locally the stresses all along the granular column thus directly confirms the Janssen description, \textit{i.e.} the exponential saturation of the stresses (Eq. 1), while previous studies inferred the validity of the model from the evolution of the apparent mass at the bottom of the column $M_a(M_g)$ (Eq. 2). 

\subsubsection{Influence of beads diameter}
\begin{figure}[h]
\centering
  \includegraphics[height=5.9cm]{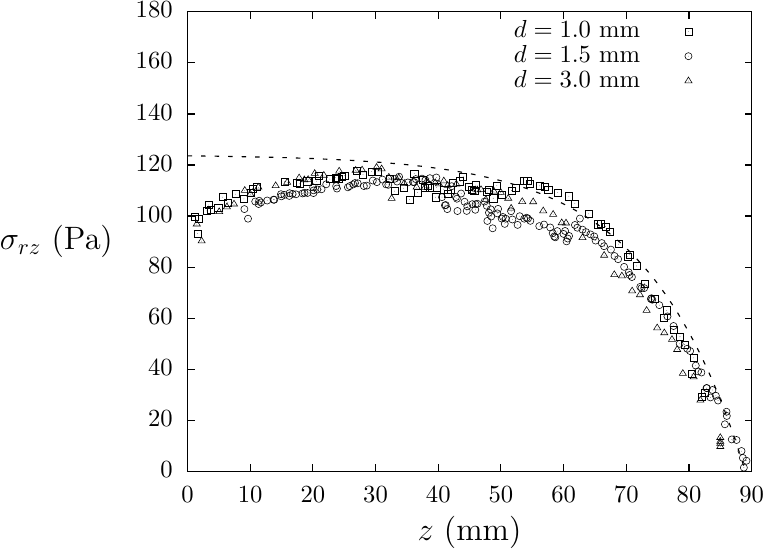}
  \caption{Quasi-static experiment: Shear stress $\sigma_{rz}$ along the tube ($M_g = 140\,$ g) with three different bead diameters ($d=1$ mm, 1.5 mm or 3 mm). The optimization procedure is repeated a hundred times for each experiment with $n=5$ estimation points of the shear stress. The dashed line corresponds to a fit of the experimental data by the function  $f(z)=[\mu_s/(1+2\mu_s^2)][gM_a/\pi R^2][1-\exp(K_w(z-l_g)/R)]$ (Eq.~\ref{rahmoun}), with $K_w=1.16$.}
  \label{fig6}
\end{figure}

We measure the stress field for the same filling length and three different bead diameters (Fig.~\ref{fig6}). We do not see any influence of bead diameter on the distribution of local stresses, as expected in the limit $d \ll R$ \cite{bratberg,qadir} for quasi-static experiments. One can note that, near the piston ($z\in[0 - 10]$ mm), the stresses are slightly smaller than the saturation value, due to the boundary effect, as it was reported numerically \cite{ovarlez2}.

\subsubsection{Effect of an overload}
\begin{figure}[h]
\centering
  \includegraphics[height=5.9cm]{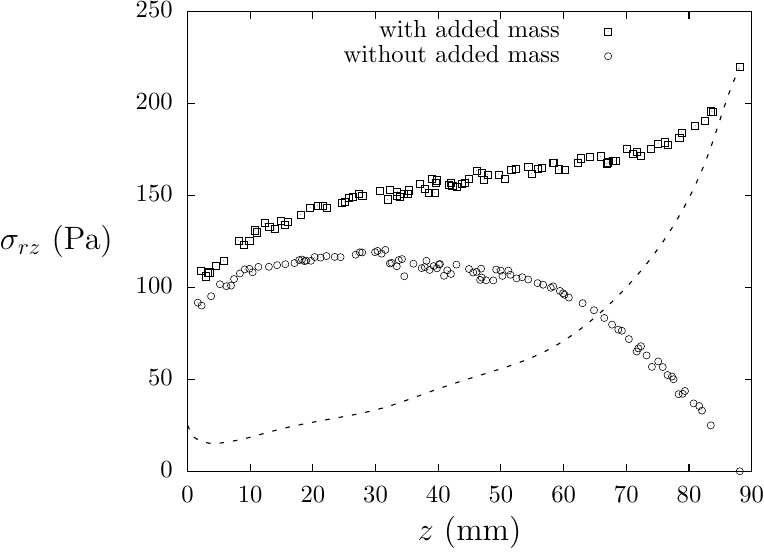}
  \caption{Quasi-static experiment: Shear stress $\sigma_{rz}$ along the tube for $M_g = 140\,$ g. Circles corresponds to the reference experiment, squares to the same experiment with an overload of $M_{ov} = 60$ g. The dashed line shows the difference between the two curves.}
  \label{fig7}
\end{figure}

Adding a weight on top of a granular media does not change the forces at the bottom of the pile as noted by Huber-Burnand. This effect was studied quantitatively in the Janssen column \cite{ovarlez1, Wambaugh2010} where the additional weight is redistributed by friction to the wall. A steel cylinder of radius slightly smaller than the radius of the tube and mass $M_{ov}=60$ g is added on top of a granular column of mass $M_g = 140\,$ g previously fully mobilized. Once the overload is added, the piston is slightly moved downwards until reaching saturation again. The stress distribution with and without the overload are plotted in Fig.~\ref{fig7}. The shear stress $\sigma_{rz}$ is maximum at the top of the column and decreases with depth to the same value than whitout the overload.  The additional shear stress at the wall, plotted in dashed line, decreases from the free surface to the bottom of the granular column. Most of the weight of the overload is screened after a typical distance of the order of the tube diameter, which is again consistent with Janssen's description.

The stress distributions in the quasi-static experiments confirm the validity of Janssen's description for confined granular materials. They complement another experimental study of the evolution of stresses in a 2D silo with photoelastic beads \cite{Wambaugh2010}. The fluctuations reported in such experiments, showing force networks in the silo, are in our case averaged close to the wall by the size of the cross-correlation windows between two pictures. However, the mean-field approach we developed could be applied to higher resolution pictures on smaller areas of the tube to track local fluctuations of stresses in the granular material at the wall. The maximum resolution would be limited by the size of the cross-correlation window which depends on the typical scale of the sputtering pattern, but also by the elasticity of the tube which imposes a typical scale of the order of the thickness.

\subsection{Local stresses during a discharge}

\begin{figure}[h]
\centering
  \includegraphics[height=5.9cm]{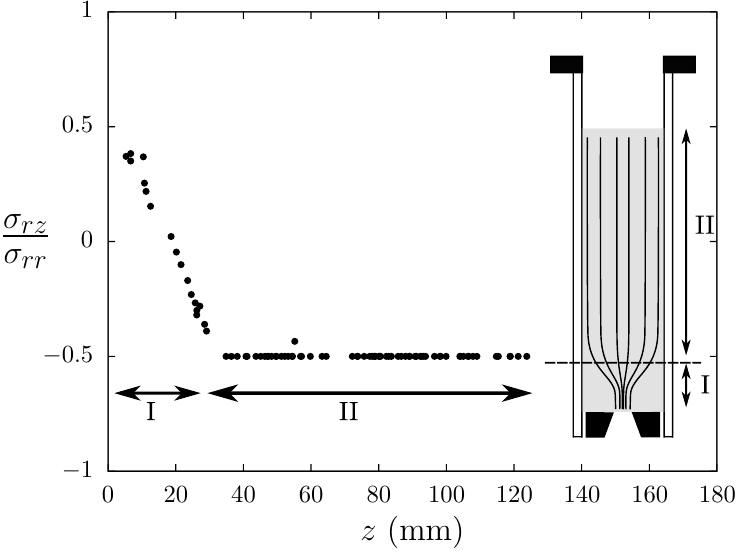}
  \caption{Discharge experiment: Stress ratio $\mu=\sigma_{rz} / \sigma_{rr}$ along the tube at $\Delta t = 0.2\ \mathrm{s}$ after the release of the plug. The grains move downwards at the wall (region II) except close to the bottom where there is no relative motion of the grains with the wall (region I). Inset : The streamlines go towards the interior of the tube \protect\cite{perge}.}
  \label{fig9}
\end{figure}

We use the same technique to determine the stress field along the silo during a gravity driven discharge. The flow rate is found to be constant ($Q=22\ \mathrm{g.s^{-1}}$) in agreement with Berverloo's law. In this experiment, after rain filling the silo, we do not mobilize friction, as in the quasi-static experiment, but directly remove the plug. In this case, the friction state is unknown and we evaluate the shear stresses and wall pressure separately. At the top, the tube is clamped and at the bottom the tube is free to move while the drilled cylinder in the outlet plane is fixed (Inset Fig.~\ref{fig9}). We measure the displacement field as a function of time and apply the optimization method to a subset of two images~: the reference image when the tube is empty and another at a given time $\Delta t$ after the plug was removed. We obtain independently the shear stress $\sigma_{rz}$ and wall pressure $\sigma_{rr}$ along the granular column for each time.

\begin{figure}[H]
\centering
  \includegraphics[height=5.9cm]{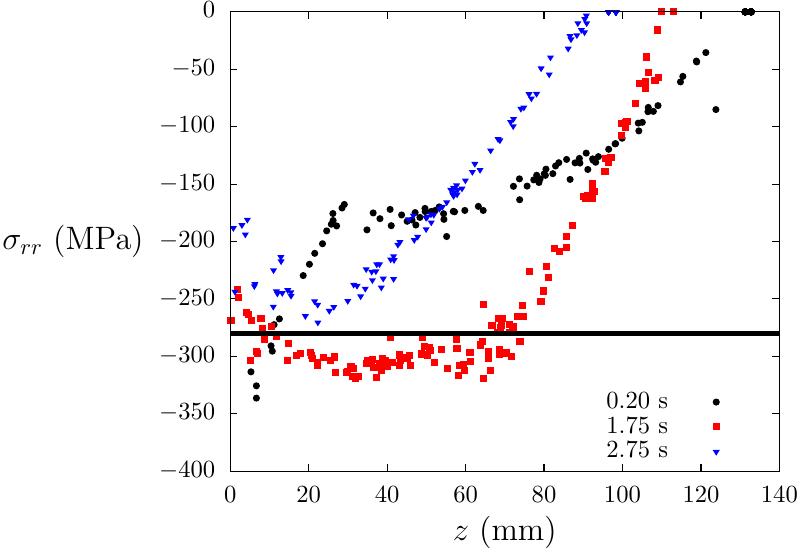}
  \caption{(Color online) Discharge experiment: Evolution of the radial stress $\sigma_{rr}$ along the tube during discharge for $M_a=220$ g. Each curve corresponds to a different time after the release of the plug, $\Delta t=0.2$ s (black circles), 1.75 s (red squares), 2.75 s (blue inversed triangles). The horizontal solid line indicates the value of the saturation for quasi-static experiments.
}
  \label{fig8}
\end{figure}

As the friction is not mobilized before the discharge, we first need to measure the stress ratio $\sigma_{rz}/\sigma_{rr}$ along the tube. This ratio is plotted just after the plug is removed (Fig.~\ref{fig9} for $\Delta t = 0.2$~s). In an upper region $z \in [30-120]$ mm (region II), the grains slide downwards at the wall ($\sigma_{rz}/\sigma_{rr} = -\mu$ (Eq.~\ref{eqvidange})), while in the lower region $z \in [0-30]$ mm (region I), there is no relative motion of the grains on the silo's wall. As the bottom part of the tube is not yet moving at this time, this indicates that the grains stand on the fixed cylinder in a still region, which is consistent with the observations of previous discharge experiments \cite{perge}: the streamlines go towards the center of the granular column close to the outlet plane (as sketched in the Inset Fig.~\ref{fig9}). The length of the region I is of the order of the tube radius. A displacement of a few grains diameters ($h=(Q\Delta t)/(\rho \phi S) \approx 10$ mm) is enough to mobilize friction on the whole column. At larger time ($\Delta t > 1$~s), as the tube and the grains move simultaneously, the stress ratio is more difficult to interpret.\\

The radial stress along the tube $\sigma_{rr}(z)$ is plotted on Fig.~\ref{fig8}. At short time ($\Delta t=0.2\ \mathrm{s}$), only a part of the grain mass is screened by the friction at the wall, and the pressure $p = -\sigma_{rr}$ saturates at a lower value than the one observed in the quasi-static experiment. Close to the fixed cylinder at the bottom ($z \in [0-30]$ mm), grains can not slide on the wall, and the pressure increases. At larger times, the friction is fully mobilized along the tube and we recover a profile caracteristic of Janssen's saturation as in the quasi-static experiment, including the pressure decrease at the bottom, due to a boundary effect. For $\Delta t = 1.75\ \mathrm{s}$, the pressure $p = -\sigma_{rr}$ increases exponentially along the tube and saturates at the value obtained in the quasi-static experiment. This description remains valid during the rest of the discharge, as the free surface of the granular column moves downwards. At $\Delta t = 2.75\ \mathrm{s}$, the grain mass in the column decreases significantly and the saturating pressure is therefore smaller.

\begin{figure}[h]
\centering
  \includegraphics[height=5.9cm]{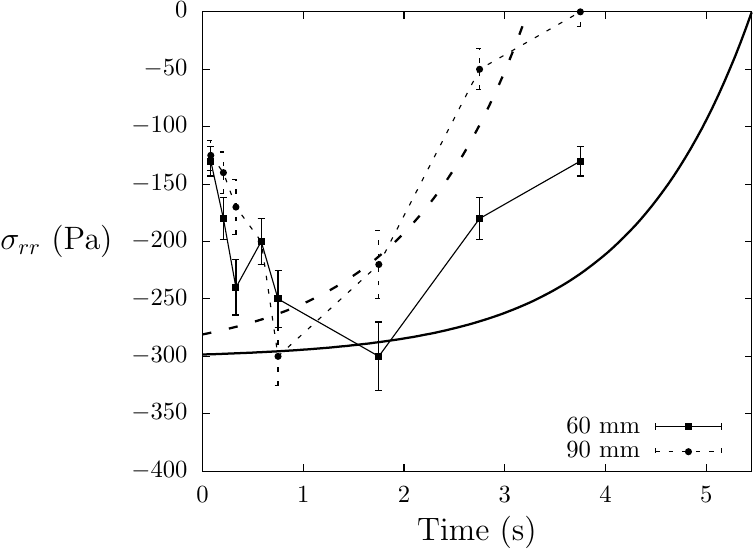}
  \caption{Discharge experiment: Evolution of the radial stress $\sigma_{rr}$ in the granular material close to the wall in time at two fixed position on the tube, $z=60$ mm (squares) and $z=90$ mm (circles). As $p = -\sigma_{rr}$, the pressure increases, reaches the saturation value corresponding to the full mobilization of friction and decreases during the discharge of the silo. The errorbars are given by the scatter of the numerical results. The solid and dashed lines correspond to the assumption that the pressure profile along the tube during all the discharge follows Janssen's law with $K_w=1.16$ obtained in the quasi-static experiments (Fig.~\ref{fig5}).}
  \label{fig10}
\end{figure}

The evolution of the radial stress at two different fixed positions along the tube (z=60 mm and 90 mm) is plotted on Fig.~\ref{fig10}. The pressure acting on the wall starts to increase as $p = -\sigma_{rr}$ until reaching the saturation value $\left|\sigma_{rr} \right| \approx 280\ \mathrm{Pa}$, obtained in the quasi-static experiments. It then decreases to zero while the tube is discharging. More precisely, if we assume that the pressure profile follows Janssen's law, with $K_w=1.16$ as in the quasi-static experiments (Fig.~\ref{fig5}), we can compute the evolution of the pressure at a given point along the tube :
\begin{equation}
	\sigma_{rr} (z,t)= -\frac{\rho \phi g R}{2 \mu_s} \left[1 - \exp{\left( \frac{K_w}{R} (z-l_g(t))\right)} \right]
\end{equation}
with $l_g(t)=l_g(t=0)-(Qt)/(\pi R^2 \rho \phi)$. The experimental measurements are in good agreement with this description (solid and dashed lines on Fig.~\ref{fig10}) for $\Delta t \ge 0.75\ \mathrm{s}$, typical time after which friction is fully mobilized. For $\Delta t < 0.75\ \mathrm{s}$, the Janssen's description is not relevant as the friction is undetermined. The typical timescale for full mobilization should depend on the radius of the tube and on the outlet diameter and is found to be very short in our experiments.

These experimental measurements of the stresses in a granular material during a gravity driven discharge show that soon after the beginning of the discharge, the stress profile close to the container's wall follows Janssen's law (except close to the outlet plane) almost until the end of the discharge.  

\section{Conclusion}
The stress repartition in a confined granular column is classically infered from the saturation of the apparent mass at the bottom of the silo. 
We measured the apparent mass as a function of filling mass in soft elastomer containers after full mobilisation of friction and recovered a unique profile described by Janssen's theory.
We show that this experimental evidence is not compatible with an isotropic elasticity description for a granular column.\\

We developed an experimental technique to measure the local stress field at the wall of a granular material confined in a vertical container.  
We validated this method against indirect measurement of the stress distribution in a granular column in a quasi-static experiment. 
The local stress distribution is found to be in agreement with Janssen's law. The beads diameter was found to have no influence on the stress profiles. 
The local distribution of additional stresses in the presence of an overload was determined.\\

Inferring local informations in a granular material from the resulting displacements of the container is a new method which could be used in a wide range of granular problems, to measure stresses under sandpiles, to track local stresses fluctuations in a granular column, or to evaluate the stress distribution in a Couette granular experiment. It could naturally be extended to study complex fluids rheology. The method developped could be applied to study other mechanical problems, as for example plant growth under constraints \cite{kolb}.

We measured the evolution of stresses in time for gravity driven discharge experiments. We found that, starting from an unknown friction state, a displacement of a few grain diameters is enough to mobilize friction on the whole column. After a short transient the pressure is found to be the same as the one observed in the quasi-static experiment. Eventually, we showed that the stress evolution in time is well described assuming Janssen's exponential saturation and Beverloo's constant mass flow rate.
\\

\textbf{Acknowledgements}
We thank E. Kolb for the lend of experimental devices, G. Lagubeau for initial discussion, J. Bico, P. Jop, E. Reyssat and L. Pugnaloni for careful proofreading of the draft.

\bibliographystyle{apsrev4-1}
\bibliography{spebib}

\end{document}